\documentclass[aps,a4paper,dvipdfmx,twocolumn,superscriptaddress,preprintnumbers,floatfix,showpacs,showkeys,letter]{revtex4}
\usepackage[dvipdfmx]{graphicx}
\usepackage[dvipdfmx]{color}
\usepackage{amscd}
\usepackage{amsmath}
\usepackage{amssymb}

\usepackage{color}

\begin{document}
\bibliographystyle{apsrev}

\title{Dynamical characterization of stochastic bifurcations in a random logistic map}

\author{Yuzuru Sato}  
\email{ysato@math.sci.hokudai.ac.jp}
\affiliation{Research Institite for Electronic Science / Department of Mathematics, Hokkaido University,  Kita 20 Nishi 10, Kita-ku, Sapporo, Hokkaido 001-0020, Japan} 
\affiliation{London Mathematical Laboratory, 8 Margravine Gardens, Hammersmith, London W6 8RH, UK}

\author{Thai Son Doan}
\affiliation{Institute of Mathematics, Vietnam Academy of Science and Technology, 18 Hoang Quoc Viet, Cau Giay, 10307 Ha Noi, Viet Nam}

\author{Jeroen S.W. Lamb}
\affiliation{Department of Mathematics, Imperial College London, 
London SW7 2AZ, UK}

\author{Martin Rasmussen}  
\affiliation{Department of Mathematics, Imperial College London, 
London SW7 2AZ, UK}


\begin{abstract}
The emergence of noise-induced chaos in a random logistic map with bounded noise is understood as a two-step process consisting of a topological bifurcation 
flagged by a zero-crossing point of the supremum of the dichotomy spectrum and a subsequent dynamical bifurcation to a random strange attractor flagged by a zero crossing point of the Lyapunov exponent. The associated three consecutive dynamical phases are characterized as a random periodic attractor, a random point attractor, and a random strange attractor, respectively. The first phase has a negative dichotomy spectrum reflecting uniform attraction to the random periodic attractor. The second phase no longer has a negative dichotomy spectrum - and the random point attractor is not uniformly attractive - but it retains a negative Lyapunov exponent reflecting the aggregate asymptotic contractive behaviour. For practical purposes, the extrema of the dichotomy spectrum equal that of the support of the spectrum of the finite-time Lyapunov exponents. We present detailed numerical results from various dynamical viewpoints, illustrating the dynamical characterisation of the three different phases. 
\end{abstract}



\preprint{arxiv.org/abs/nlin/XXX}

\maketitle

Noise-induced phenomena arise out of the interaction between nonlinear and stochastic terms in random dynamical systems. Stochastic resonance \cite{benzi1982stochastic}, noise-induced synchronization \cite{pikovskii1984synchronization},
\cite{teramae2004robustness}
  and noise-induced chaos \cite{mayer1981influence} are representative examples of noise-induced transitions in 
nonlinear systems that may be observed as a result of a change of the noise level \cite{crutchfield1982fluctuations}. 

Despite of the importance of such transitions in applications, the theory of such phenomena is still not well developed with a number of potential problems in sciences still remain unsolved. 
The theory of
random dynamical systems addresses such problems, with the potential to  extend 
nonlinear physics far beyond the traditional scope. Recently, novel insights into bifurcations of random dynamical systems have been obtained in \cite{zmarrou2007bifurcations, lamb2011topological, engel2016bifurcation, callaway2017dichotomy}, providing a more dynamical characterization, 
taking into account pathwise dynamical behaviour beyond the information 
from aggregate measures such as the stationary distribution. Moreover, the consideration of bounded noise also opens up the possibility to address 
topological changes of random attractors. In this letter, we build on these novel insights to detail dynamical characterisations of noise-induced transitions. We focus on a logistic map with additive noise at a parameter value where
the underlying deterministic map has a unique attracting period three point (and many unstable periodic points). 
As a function of increasing noise we observe three phases of random dynamics, corresponding a random periodic attractor, a random fixed point attractor and random strange attractor, with transition points characterized by zero-crossing points of the supremum of the dichotomy spectrum and Lyapunov exponent, respectively. 

We consider the logistic map with additive noise
\begin{equation}
  x_{n+1}=ax_n(1-x_n)+\omega_n,
\label{eq:RLM}
\end{equation}
where $n=0,1,2,\ldots$. The second term $\omega_n\in[-\epsilon,+\epsilon]$ 
is a random variable chosen i.i.d. with uniform distribution. 
It has been numerically observed \cite{mayer1981influence} that when $a=3.83$ and the deterministic logistic map has an attractive period three orbit, the random logistic map exhibits a noise-induced chaotic attractor, evidenced by a broadband power spectrum, a positive Lyapunov exponent, and positive Kolmogorov-Sinai entropy \cite{crutchfield1982fluctuations}. This chaotic attractor emerges due to the interaction between the stable periodic orbit and the unstable chaotic repeller, enabled by the noise \cite{mayer1981influence, crutchfield1982fluctuations}. 
This transition from a non-chaotic to a chaotic attractor due to noise, is understood to arise when noise is added to a deterministic system with an attractor with negative Lyapunov exponent and positive topological entropy. 

In the literature \cite{mayer1981influence}, the critical noise amplitude at which the chaotic attractor appears, is known as $\epsilon_L\approx 0.00146$, when the Lyapunov exponent passes through zero from below with increasing noise amplitude (and turns positive with larger noise amplitude). In \cite{tel2010quasipotential}, scaling of the Lyapunov exponent (as an order parameter) has been considered as a means to characterize this transition. 

Here, we provide a more detailed bifurcation analysis for the build-up towards noise-induced chaos in this regime of the random logistic map (\ref{eq:RLM}). 
In particular, we identify a topological bifurcation point 
at $\epsilon=\epsilon_D\approx 0.000251$ where the support of the stationary measure explodes \cite{mayer1981influence,lamb2011topological}. This is accompanied by dynamical changes in terms of the supremum of the {\it dichotomy spectrum} traversing through $0$ and a change from uniform attractivity of the random point attractor to nonuniform attractivity.

We thus obtain transition points that divide the $\varepsilon$-parameter space into three complementary regions: phase I ($0<\epsilon<\epsilon_D$), phase II ($\epsilon_D<\epsilon<\epsilon_L$), and phase III ($\epsilon_L<\epsilon$). 
\begin{figure}[htbp]
    \centering
	\includegraphics[bb=0 0 411 568, scale=0.55]{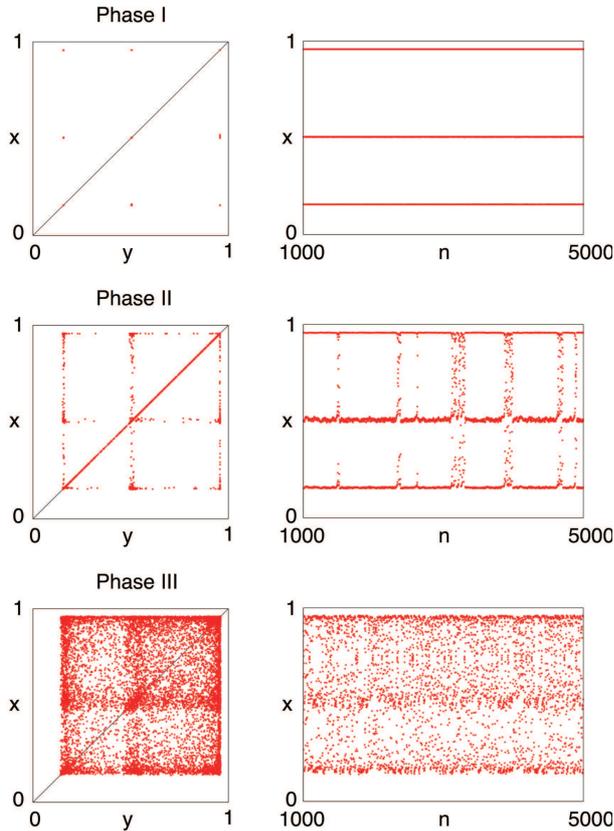}
\caption{Examples of two point motion diagram and time series in the phase I (top, $\epsilon=0.0001$, random period-three attractor), the phase II (middle, $\epsilon=0.001$, random point attractor with synchronization), and the phase III (bottom, $\epsilon=0.004$, random strange attractor).}
\label{fig:tpp}
\end{figure}
\begin{table}[htbp]
  \begin{tabular}{|c|c|c|c|c|}\hline
    phase & attractor & $\sup\Sigma$ &  ~$\lambda$~ &  $\#$cs \\\hline\hline
    I &random periodic attractor  & $-$ & $-$ &  3 \\\hline
    II &random point attractor &$+$ & $-$ & 1 \\\hline
    III &random strange attractor&$+$ &  $+$ &  1 \\\hline
  \end{tabular}
\caption{A table of properties of phase I-III. The labels $\sup\Sigma$, $\lambda$, $\#$cs mean the supremum of dichotomy spectrum, Lyapunov exponent, and the number of connected components of the support of the stationary measure, respectively.}
\label{table:phase}
\end{table}

In the phase I, we observe a random point attractor with periodic motion. The support of the invariant density for this attractor consists of three mutually disjoint intervals. 
The random attractor is a uniformly attracting random period three point, consisting of three random points $\{p_1(\omega),p_2(\omega),p_3(\omega)\}$, each one assigned exclusively to one of the aforementioned intervals, that are being cyclically permuted by the dynamics. As the noise amplitude passes $\varepsilon_D$, in phase II, the support of the stationary measure explodes to a connected set. The Lyapunov exponent remains negative, but the attractor turns into an attracting random fixed point. This means that in contrast to the well-defined periodic behaviour in phase I, where two orbits having initial conditions in different connected components of the stationary measure can never converge towards each other, in phase II the distance between any two orbits with different initial conditions eventually tends to zero for almost every realization of the noise \cite{callaway2017dichotomy}. The latter phenomenon is known as noise-induced synchronization\cite{pikovskii1984synchronization, teramae2004robustness}. During phase II, as the noise level increases, the density of the stationary measure gradually becomes more uniformly spread. 

In the transition from phase II to phase III, the Lyapunov exponent turns positive and the random fixed point attractor bifurcates to a random strange attractor. Whereas the attracting random fixed point supports synchronization, on the random strange attractor orbits of different initial conditions typically do not converge to each other. 

The changes in dynamical behaviour between the three phases are well viewed through the {\it two point motion} in which we follow a pair of initial conditions $\{x_0,y_0\}$ for the same one-dimensional dynamical system: 
\begin{equation}
\left\{\begin{array}{rcl}
x_{n+1}&=&ax_n(1-x_n)+\omega_n, 
\\
y_{n+1}&=&ay_n(1-y_n)+\omega_n, 
\label{eq:tpp}
\end{array}\right. 
\end{equation}
The two point motion thus consists two logistic maps coupled through common noise. In phase I, we have $9$ clusters corresponding to the support of the three distinct random period three attractors (FIG. \ref{fig:tpp}). In phase II, we observe a strong concentration of density near to the diagonal $x=y$, representing synchronization of $x$ and $y$. In phase III, we have chaotic dynamics and no longer observe synchronisation. Properties of these three phases are summarized in TABLE I. Phase II is a new characterization of dynamics which has been overlooked previously in studies of noise-induced chaos. 
Due to the negative Lyapunov exponent, the support of the invariant density is a subinterval of the diagonal $x=y$. Asymptotically, orbits converge almost surely to this diagonal although each nearly synchronized state (an initial condition near the diagonal) has a positive chance to desynchronize (move away from the diagonal). We observe this as chaotic desynchronisation bursts in this model. These chaotic bursts do not contribute to asymptotic statistics and this is reflected in the fact that the stationary measure sits on the diagonal. The dynamics of chaotic bursts in phase II with negative Lyapunov exponent, and impermanent trapping near periodic motion in phase III with positive Lyapunov exponent are nontrivial and a more detailed analysis will be reported elsewhere. With phase II as a generic precursor to phase III we hereby identify a universal route to stochastic chaos in random dynamical systems. 

To define Lyapunov exponent and dichotomy spectrum for the random attractors in the random logistic map, we briefly formalize the setting. We may write Eq. (\ref{eq:RLM}) as 
\begin{equation}
  x_{n+1}=\phi(1,\theta^n\omega, x_n)
\end{equation}
where $\omega=\{\omega_n\}_{n\in\mathbb Z}\in \Omega:=[-\epsilon,+\epsilon]^{\mathbb Z}$.
Here $\theta$ represents the evolution of the noise. 
While the whole dynamics $(\theta, \phi)$ works in the extended space $\Omega\times \mathbb{R}$, we are interested in the projected dynamics in $\mathbb{R}$.  A random pullback attractor $A(\omega)\subset \mathbb{R}$ satisfies 
\begin{equation}
\lim_{n\rightarrow\infty} \mbox{dist}(\phi(n,\theta^{-n}\omega,B), A(\omega))=0, 
\end{equation}
for any bounded set $B$, where dist is the Hausdorff distance \cite{arnold1998random}. 
The Lyapunov exponent $\lambda(\omega,x_0)$ is in principle a function of the noise realization $\omega$ and the initial condition $x_0$. However, here it is almost surely constant $\lambda(\omega, x_0)=\lambda$
\cite{viana2014lectures}. 
The Lyapunov exponent for Eq. (\ref{eq:RLM}) is given as 
\begin{equation}
\lambda=\lim_{N\rightarrow\infty}\frac1N\sum_{n=0}^{N}\ln|(a(1-2x_n)|.
\end{equation}
The sign of $\lambda$ indicates local stability of the dynamics with typical noise realization. A random attractor is called a random strange attractor, when the associated Lyapunov exponent is positive \cite{lin2009reliability,chekroun2011stochastic, faranda2017stochastic}. In this case, trajectories of nearby initial conditions on the attractor tend to separate from each other asymptotically. The associated chaotic behavior observed in such stochastic dynamics is called stochastic chaos \cite{kapitaniak1986stochastic,arnold1988lyapunov}. In contrast, if the Lyapunov exponent is negative, trajectories of nearby initial conditions tend to synchronize. 


The dichotomy spectrum is a measure of uniform expansion and contraction in the dynamics. It turns out that the range of its support coincides with the observability range of finite-time Lyapunov exponents in the infinite-time limit \cite{callaway2017dichotomy}.
The finite-time Lyapunov exponents of the random attractor $A(\omega)$ are defined by 

\begin{equation}
  \lambda(N,\omega,x_0)=\frac1N\sum_{n=0}^{N}\ln|(a(1-2x_n)|, ~(x_0\in A(\omega)).
\end{equation}
The Lyapunov exponent is given by $\lim_{N\rightarrow\infty}\lambda(N, \omega, x_0)$. Note that, in contrast to the Lyapunov exponent, a finite-time Lyapunov exponent typically depends on the initial conditions. The standard deviation of finite-time Lyapunov exponent converges to $0$ when the Lyapunov exponent exists.

The dichotomy spectrum $\Sigma$ of Eq. (\ref{eq:RLM}) is the complement of the set of growth rates $\gamma_-,\gamma_+\in\mathbb R$ that satisfy for some  $K_{-}, K_{+} >0$ 
\begin{equation}
K_{-} e^{(\gamma_{-}-\alpha)n}\le|D_x\phi(n,\omega,x)|\le K_{+} e^{(\gamma_{+}+\alpha)n}, 
\label{eq: exdic2}
\end{equation}
for almost all $\omega$, all $n,\alpha >0$, and all $x$: 
\begin{eqnarray}
  \Sigma=\mathbb{R}\setminus (\{\gamma_{-}\}\cup\{\gamma_{+}\}).
\end{eqnarray}

The relationship between the supremum/infimum of the dichotomy spectrum and the finite-time Lyapunov exponent is given by \cite{callaway2017dichotomy}
\begin{eqnarray}
  &&\sup\Sigma=\lim_{N\rightarrow\infty}\mbox{ess}\sup_{\omega}\sup_{x_0}\lambda(N, \omega, x_0),\nonumber\\
   &&\inf\Sigma=\lim_{N\rightarrow\infty}\mbox{ess}\inf_{\omega}\inf_{x_0}\lambda(N, \omega, x_0), 
\end{eqnarray}
where the ess sup is the supremum except for null measurable sets. As it turns out that in the one-dimensional setting the support of the dichotomy spectrum is connected, in practice, it suffices to determine these upper and lower bounds.

\begin{figure}[htbp] 
  \centering
	\includegraphics[bb=0 0 340 338, scale=0.35]{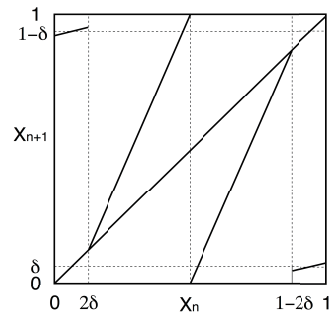}
\caption{Random piecewise-linear map showing noise-induced chaos. In the deterministic limit, the dynamics of Eq. (\ref{eq:ex1}) shows chaotic transients converging to a stable period-two limit cycle with the periodic points $x=\delta$ and $x=1-\delta$. Noise-induced chaotic burst can be observed when $\epsilon > \delta (1-\kappa)$. }
\label{fig:ex}
\end{figure}

To illustrate the benefit of the dichotomy spectrum, we give an example of a random piecewise-linear map
$x_{n+1}=h(x_n)+\omega_n$, which shows noise-induced chaos:  
\begin{equation}
h(x)=\left\{
\begin{array}{ll}
\kappa(x-\delta)+1-\delta& (x < 2\delta)\\
L(x-2\delta)+2\delta&(2\delta\le x < \frac12)\\
L(x-\frac12)&(\frac12\le x\le 1-2\delta)\\
\kappa(x-1+\delta)+\delta & (1-2\delta< x)
\end{array}
\right.,
\label{eq:ex1}
\end{equation}
where $L=\frac{2-4\delta}{1-4\delta}$, $\delta\in(0,1/2)$, $\kappa\in[0, 1)$, and the random variable $\omega_n\in[-\epsilon,+\epsilon]$ follows uniform distribution (See FIG. \ref{fig:ex}).

One can find that the dynamics of Eq. (\ref{eq:ex1}) shows chaotic transients converging to a stable period-two limit cycle in the deterministic limit. When $\epsilon > \delta(1-\kappa)$, the orbit in the attractor may be kicked out from the uniformly attracting region $[-\infty,2\delta)\cup(1-2\delta,+\infty]$, and chaotic bursts are observed. The Lyapunov exponent is given as a monotonically increasing continuous function of $\epsilon$ greater than $\ln\kappa$. The zero-crossing point of the Lyapunov exponent characterize the transition point of stability of the random attractor. On the other hand, the dichotomy spectrum is given by the following discontinuous function of $\epsilon$:
\begin{equation}
\Sigma=\left\{
\begin{array}{ll}
\{\ln \kappa \} & (\epsilon < \delta(1-\kappa))\\
\left[\ln \kappa, ~\ln L\right]& (\epsilon \ge \delta(1-\kappa))
\end{array}
\right..
\label{eq:dsex}
\end{equation}
Thus, the zero-crossing point of the supremum of the dichotomy spectrum, $\epsilon=\delta(1-\kappa)$, gives the topological bifurcation point to noise-induced chaos.

For the random logistic map given by Eq. (\ref{eq:RLM}), when the period 3 limit cycle is before the bifurcation to the super-stable cycle at $a_c\approx 3.831874055283321$, the noise amplitude $\epsilon_D$ with which the dichotomy spectrum contains 0, is given as the smallest real $\epsilon$ solving the following equations; 
\begin{equation}
g_k(g_j(g_i(x)))=x, ~\left|\frac{\partial}{\partial x}[g_k(g_j(g_i(x)))]\right|=1, 
\end{equation}
where $i,j,k\in\{0,1\}$, $g_0(x)=ax(1-x)-\epsilon$, $g_1(x)=ax(1-x)+\epsilon$. 
If the critical point is contained in the support of the stationary measure, 
the dichotomy spectrum is not bounded from below. 
Our numerical computation suggests that the supremum of dichotomy spectrum of
the random logistic map with $a=3.83$ is $0$ when $\epsilon_D\approx 0.000251396$. 

\vspace{2mm}
Based on the above investigation, we give an algorithm measuring the supremum of the dichotomy spectrum of Eq. (\ref{eq:RLM}). Assuming that the dichotomy spectrum of the random dynamical systems $(\theta, \phi)$ is given by $\Sigma=[b,c]$, we can estimate $c$ based on a rare orbit which has the maximal finite-time Lyapunov exponent in the time interval $[0,N]$. One may adopt a certain type of Monte Carlo algorithm to find such a rare orbit \cite{tailleur2007probing}. For Eq. (\ref{eq:RLM}) with $a=3.83$, there is a local search algorithm which works more efficiently than exhaustive Monte Carlo search. In this case, local optimization of the largest finite-time Lyapunov exponents works because the derivative of Eq. (\ref{eq:RLM}) is linear and the trajectories converge uni-directionally to the period-three limit cycle. Starting with an initial condition $x_0$ in the random attractor, at each step $n$, we generate $g_0(x_n)$ and $g_1(x_n)$ as candidates of the rare orbit for the next state, evaluate the finite-time Lyapunov exponents, and repeatedly renew it to the larger one. In a long run with a large set of initial condition in the random attractor, it converges to $c$. We can compute $b$ similarly, but it does not have particular importance here. The value of $\epsilon$ for which $c=0$ corresponds to the topological bifurcation point $\epsilon_D$. Note that if we practically find one example of $c>0$, it implies that there exists an orbit which gives a positive finite-time Lyapunov exponent with a positive probability. After $\epsilon>\epsilon_D$, the numerical estimation  of $c$ is more complicated, however the sign of $c$ is easy to obtain and more important than the value for our purpose. 
\begin{figure}[htbp]
    \centering
    \includegraphics[bb=0 0 403 385, scale=0.6]{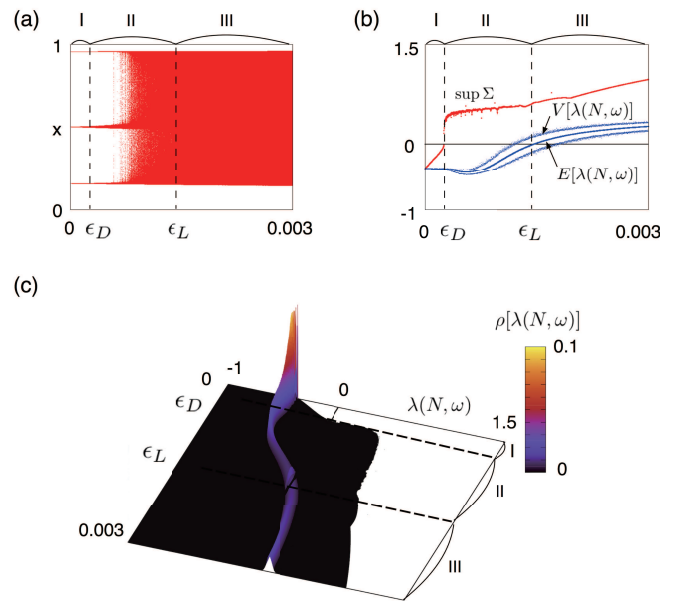}
\caption{(a) Typical orbits vs. noise amplitude $\epsilon$ in random logistic maps, (b) Diagram of Lyapunov exponents and supremum of dichotomy spectrum as a function of noise amplitude $\epsilon$, including the average of finite-time Lyapunov exponent $E[\lambda(N,\omega)]$, which approximates Lyapunov exponent, 
  the variance of finite-time Lyapunov exponent $V[\lambda(N,\omega)]$, and the supremum of dichotomy spectrum $\sup\Sigma$. (c) The distribution of the finite-time Lyapunov exponents $\rho [\lambda(N,\omega)]$. The black region indicates finite positive density, where the finite-time Lyapunov exponents are observable, while the white region indicates no chance to observe those. All figures are computed with $a=3.83$, $\epsilon \in[0,0.003]$, and $N=3000$. }
\label{fig:bif}
\end{figure}

The FIG. \ref{fig:bif}(a) is the transition diagram from period $3$ limit cycle to chaos when we increase the noise amplitude $\epsilon$. 
In the FIG. \ref{fig:bif}(b), we depict the numerical estimation of the average of finite-time Lyapunov exponent $E[\lambda(N,\omega)]=\langle \lambda(N, \omega) \rangle_{\omega}$, the variance of finite-time Lyapunov exponent $V[\lambda(N,\omega)]=\langle \lambda(N, \omega)^2\rangle_{\omega}-\langle \lambda(N, \omega)\rangle_{\omega}^2$, and the supremum of dichotomy spectrum $\sup\Sigma$, for Eq. (\ref{eq:RLM}) with $a=3.83$, $\epsilon \in [0,0.003]$, and $N=3000$ (FIG. \ref{fig:bif}(b)). FIG. \ref{fig:bif}(c) gives three-dimensional view of the distribution of the finite-time Lyapunov exponents.


The black region indicates the support of the finite-time Lyapunov exponent spectrum with small probability of observation.
The white region lies outside the support of the finite-time Lyapunov exponent spectrum. 



To summarize, in the random logistic map, we identify two transition points related to the zero-crossing points of the Lyapunov exponent and the supremum of the dichotomy spectrum. As a result, the parameter space is divided into three regions characterised by a random periodic attractor, a random point attractor and a random strange attractor, respectively. The change of sign of the Lyapunov exponent implies appearance / disappearance of eventual local synchronisation of trajectories. 
The loss of uniform hyperbolicity of the dichotomy spectrum coincides with a topological bifurcation point of the random periodic attractor, which reveals dynamical properties that are not observable through the Lyapunov exponent. At a technical level, the difference between these spectral concepts concerns the required uniformity of hyperbolicity.

We note that just beyond the critical noise amplitude $\epsilon_D$, chaotic bursts are initially rare but become more frequent when the noise amplitude is further increased. The transition point is the theoretical threshold for the possibility of such bursts. It is a topological bifurcation point in random dynamical systems, and the dichotomy spectrum (or supremum of the finite-time Lyapunov exponents) can be utilized as an early warning signal \cite{scheffer2009early} for this noise-induced transition. There are a number of applications that require an absolute absence of such transitions, such as in power failure, computer security, traffic disruption, and so on. We have illustrated the use of random dynamical systems theory to provide appropriate concepts and powerful methodologies to understand nonlinear stochastic phenomena.

\section*{Acknowledgments}
YS was supported by JSPS Grant-in-Aid for Scientific Research (C) No. 24540390 and No. 18K03441 and EPSRC platform grant (EP/I019111/1), JSWL by Nizhny Novgorod University (RNF 14-41-00044) and MR by an EPSRC Career Acceleration Fellowship (EP/I004165/1). YS and JSWL thank the London Mathematical Laboratory for support. The authors were also supported by EU ITN Critical Transitions in Complex Systems (H2020-MSCA-2014-ITN 643073 CRITICS). 

\bibliography{nibrlm}


\end{document}